\documentclass[structabstract]{aa}
\usepackage{graphicx,url}
\usepackage{txfonts}
\usepackage{natbib}
\begin{document}
\title{Characteristics of polar coronal hole jets}
\author{K.~Chandrashekhar\inst{1}
          \and
          {A.~Bemporad}\inst{2}
           \and
          D.~Banerjee\inst{1}
          \and
          G.~R.~Gupta\inst{3}
          \and
          L.~Teriaca\inst{3}
          }

   \institute{Indian Institute of Astrophysics, Koramangala, Bangalore 560034\\
             \email{dipu@iiap.res.in}
             \and
             INAF - Turin Astrophysical Observatory, via Osservatorio 20, 10025 Pino Torinese(TO), ITALY
             \and
             Max-Plank-Institut f\"{u}r Sonnensystemforschung (MPS), 37191, Katlenburg-Lindau, Germany
             }

   \date{\today}

   \abstract
   {High spatial- and temporal-resolution images of coronal hole regions show a dynamical   environment where mass flows and jets are frequently observed.  These jets are believed to be important for the coronal heating and the acceleration of the fast solar wind.  } 
  {We studied the dynamics of two jets seen in a polar coronal hole with a combination of imaging from EIS and XRT onboard Hinode. We observed drift motions related to the evolution and formation of these  small-scale jets, which we  tried to model as well. }
  {Stack plots were used to find the drift and flow speeds of the jets. A toymodel was developed by assuming that the observed jet is generated by a sequence of single reconnection events where single unresolved blobs of plasma are ejected along open field lines, then expand and fall back along the same path, following a simple ballistic motion.}
  {We found observational evidence that supports the idea  that polar jets are very likely produced by multiple small-scale reconnections occurring at different times in different locations. These eject plasma blobs that flow up and down with a motion very similar to a simple ballistic motion. The associated drift speed of the first jet is estimated to be $\approx$ 27 km s$^{-1}$. The average outward speed of the first jet is $\approx 171$ km s$^{-1}$, well below the escape speed, hence if simple ballistic motion is considered, the plasma will not escape the Sun.  The second jet was observed in the south polar coronal hole with three XRT filters, namely, C$_{-}$poly, Al$_{-}$poly, and Al$_{-}$mesh filters. Many small-scale ($\approx 3\arcsec-5 \arcsec$) fast ($\approx$ 200 - 300 km s$^{-1}$) ejections of plasma were observed on the same day; they propagated outwards. We observed that the stronger jet drifted at all altitudes along the jet with the same drift speed of $\simeq$ 7~km~s$^{-1}$. }
   {The enhancement in the light curves of low-temperature EIS lines in the later phase of the jet lifetime and the shape of the jet's stack plots suggests that the jet material is falls back, and most likely cools down. To further support this conclusion,  the observed drifts were interpreted within a scenario where reconnection progressively shifts along a magnetic structure, leading to the sequential appearance of jets of about the same size and physical characteristics. On this basis, we also propose a simple qualitative model that mimics the observations.}

   \maketitle
\section{Introduction}
Polar jets and X-ray bright points are prominent dynamical features of coronal hole regions. Soft X-ray jets were discovered in Yohkoh/SXT data \citep{1992PASJ...44L.173S, 1992PASJ...44L.161S}. In the coronal hole regions the ambient magnetic field is nearly  vertical and often unipolar,  and interacting with the emerging field gives rise to reconnection followed by mass ejections, with collimated hot-plasma flows commonly termed jets  \citep{1995Natur.375...42Y}. After the launch of Hinode/XRT it was discovered that these jets occur more frequently than previously thought, at a frequency of between 60 jets per day \citep[as reported by][]{2007PASJ...59S.771S} and 10 jets per hour \citep[as reported by][]{2007Sci...318.1580C}. Small-scale solar eruptions, seen in different wavelengths are often termed differently, for instance, H$\alpha$ surges \citep{1934MNRAS..94..472N}, spicules (e.g., Secchi 1877; \citealp{1972ARA&A..10...73B} and references within), type II spicules \citep{2007PASJ...59S.655D}, macro-spicules \citep{1975ApJ...197L.133B}, UV jets \citep{1983ApJ...272..329B}, EUV jets \citep{1998A&A...334L..77B}, and X-ray jets \citep{1992PASJ...44L.173S}. Recent high-resolution and high-cadence observations (e.g., using SOHO, Hinode, STEREO, Solar Dynamic Observatory - SDO) have allowed a detailed study of coronal hole jets, providing information on their inherent dynamic behavior (e.g., \citealp{KAMetal2007}; \citealp{PATetal2008};\citealp{NISetal2009};  \citealp{FILetal2009} \citealp{LIUetal2009, LIUetal2011}; \citealp{KAMetal2010};  \citealp{SHEetal2011}; \citealp{YANetal2011};  \citealp{MORetal2012b};  \citealp{CHEetal2012}).

Jet observations acquired by the Hinode or SOHO spacecraft high-resolution imagers and spectrometers can give precise measurements of the plasma flow velocities in the coronal hole region. This information and the possible information on the magnetic topology of the coronal hole region is very important in interpreting coronal heating and solar wind acceleration. In recent times there have been claims that these jets can substantially contribute to the plume formation \citep{2008ApJ...682L.137R} and in turn can be important for the energy budget of the wind. An interesting topic related with the study of polar  coronal hole jets is the amount of mass possibly ejected in the higher corona, that may hence be able to provide a source for the fast wind. \cite{2007PASJ...59S.751C} observed post-jet enhancements in light curves corresponding to cooler lines and explained them as a signature of plasma that is at first heated to coronal temperatures and then falls back on to the Sun along open magnetic-field lines after cooling. Signatures for a cooler plasma downflow were observed in the decay phase of an EUV jet by \cite{2007PASJ...59S.751C}, \cite{2009ApJ...704.1385S}, \cite{2011A&A...526A..19M}, and more recently by \cite{2012ApJ...759...15M}, who found that the density of the cool plasma in a jet was ``larger than that of background cool plasma'', but who  were unable to explain the origin of this difference. All these works suggest that the real impact of EUV jets on the fast solar-wind flow could be very limited.

\begin{figure}
\resizebox{\hsize}{!}{\includegraphics{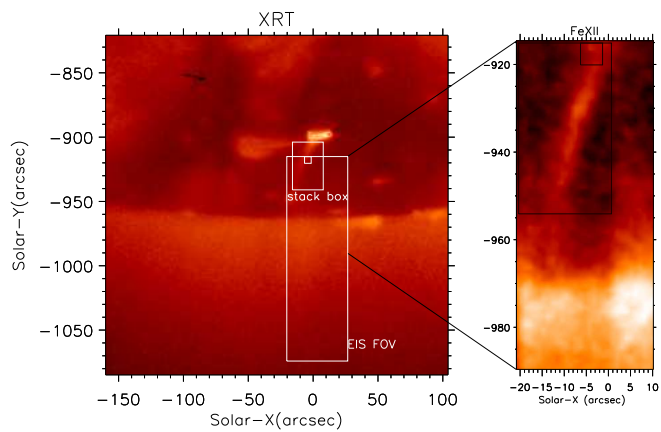}}
\caption{Left panel: XRT image taken using the thin Al$_{-}$mesh filter at 13:12 UT on April 15, 2007 (jet~1). The large rectangular box marks the  EIS FOV corresponding to the 40$\arcsec$ wide EIS slot. Right panel: correspond to an image of the jet as seen in EIS \ion{Fe}{xii}. The small square box corresponding to  the upper body of the jet is used for light curve analysis.} The intermediate size rectangular box in the left panel marks the area used to produce XRT stack plot.
\label{context}
\end{figure}
In this article we report on the evolution of two jets that have been observed in the southern polar coronal hole. Because the Sun was in  deep minimum in April 2007, we had clear coronal holes, which appear dark in X-ray images. On April 15, 2007 we observed an EUV jet on the disk in the southern polar coronal hole simultaneously with Hinode/EIS and XRT and with SOHO/CDS. A number of small bright formations that were identified as X-ray bright points were also observed. In July 2008 we observed another EUV jet off-limb above the southern coronal hole during a coordinated SOHO/SUMER and Hinode/XRT campaign. In section~\ref{obs}, we describe the data of these two jets and the different corrections we applied.  In section~\ref{res} we describe and quantify the drift motion observed in both jets and provide an interpretation for this drift motion in terms of reconnections that progressively occur in magnetic structures along the neutral line. In this section we also discuss other interesting properties of the individual jets, such as the jet flow-speed and light curves, which clearly show post-jet enhancements. In section~\ref{model} we develop a toy model to interpret the observed drift motion that is associated with the jets. It is important to note that both jets were associated with an X-ray bright point, but the first jet (April~2007) was observed entirely on disk, while the second jet (July~2008) was observed entirely off-limb. Conclusions are given in the final section.

\section{Observations}
\label{obs}
\subsection{April 2007 data}

The data analyzed here were obtained on April 15, 2007 during a Hinode/SoHO joint observing campaign as part of the  Hinode Observing Programme (HOP)~45. They consist of time series taken in the south polar coronal hole by the EUV imaging spectrometer \citep{2007PASJ...59S.751C} and the X-Ray Telescope \citep[XRT]{2007SoPh..243...63G}, both onboard Hinode \citep{2007SoPh..243....3K}. The XRT data consist of images taken using the thin Al mesh filter from 10:54 UT to 15:58 UT with an effective cadence of 18~s. For EIS, the $40\arcsec$ wide slot was used to obtain  $40\arcsec \times  160\arcsec$ images in several spectral lines over the time interval 10:54 to 15:57 UT. The EIS data consist of a series of elementary rasters formed by two slot images displaced by 20$\arcsec$ in the X direction. Each slot image was exposed for about 7.5~s. As a result, we have two time series with an effective cadence of 19.3~s, of which only one used here. The location of the EIS and XRT field of view (FOV) are shown in Figure~\ref{context}.  All the data were reduced and calibrated with the standard procedures given in the SolarSoft(SSW) library (see also the online movie of the jet in XRT, movie~1: \url{ftp://ftp.iiap.res.in/chandra/xrt-jet/movie1.mpeg} \label{m1}). The jitter effect was corrected for using the housekeeping data; the movement of the slot image on the detector due to the thermal variations during the orbit was also corrected. The EIS data have been used by \cite{2009A&A...499L..29B} for the detection of waves in the off-limb part of the coronal hole; more details of the EIS data reduction can be found there. 

For XRT, the standard reduction and calibration routines are provided by the xrt$_{-}$prep procedure. Instrumental noise and cosmic-particle hits were removed from the data. The images were normalized by exposure time.

\subsection{July  2008 data }
\begin{figure*}
\resizebox{\hsize}{!}{\includegraphics{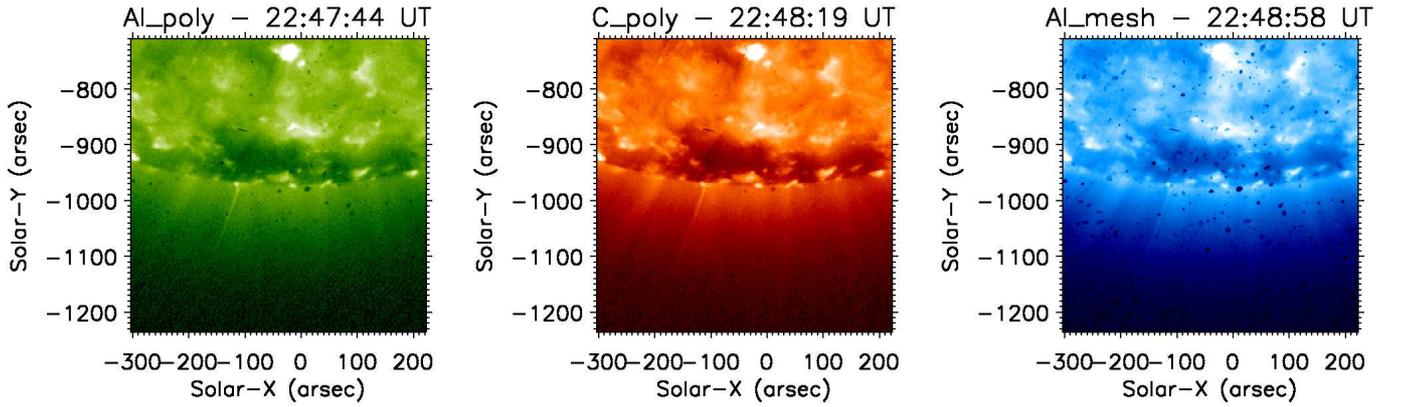}}\vspace*{-1cm}
\caption{Example of an XRT image sequence acquired on July 1, 2008 around 22:48 UT during polar-jet event~2 with the Al$_{-}$poly (left), C$_{-}$poly (middle), and Al$_{-}$mesh (right) filters (see also the movie~2).}
\label{context2}
\end{figure*}

A multi-instrument campaign involving SOHO, Hinode, and TRACE has been conducted from June 22 to July 3, 2008. Observations focused on the south polar coronal hole. Hinode/XRT observations were acquired by alternatively using the Al$_{-}$poly, C$_{-}$poly, and Al$_{-}$mesh filters, with exposure times of 23~s and a cadence of 35~s. The FOV was 512 $\times$ 512 pixels ($526.6\arcsec \times 526.6\arcsec $) centered on average at $X=-39.9\arcsec$ and $Y=-972.3\arcsec$. On July 1, 2008 we identified seven polar jets during $\sim 3$ hours of observing time. These jets, lasting for 4 to 17 minutes, are typically associated with plumes. \cite{2008ApJ...682L.137R} have also reported that 90\% of the 28 jets observed by them were associated with polar plumes. A couple of our jets originated from the same region within 10 minutes. The best-visible event started at 22:44 UT and lasted until 23:02 UT. A sequence of images in the three different filters acquired around 22:48 UT during the jet event is shown in Figure~\ref{context2}, see also the online movie~2 \url{ftp://ftp.iiap.res.in/chandra/xrt-jet/movie2.gif} \label{m2}.

\section{Results: Jet properties}
\label{res}
In this section we discuss the different characteristics of the observed jets. We selected  jets that were long-lasting in nature and show drifts. We aim to understand the dynamics of the jets in terms of their source and life time. In the following subsections we separately describe different properties such as the drift speed, the lifetime and the expansion speed, and evolution  of the associated sigmoid structure.
\subsection{Drift speed}
\label{dri}
\begin{figure}
\resizebox{\hsize}{!}{\includegraphics{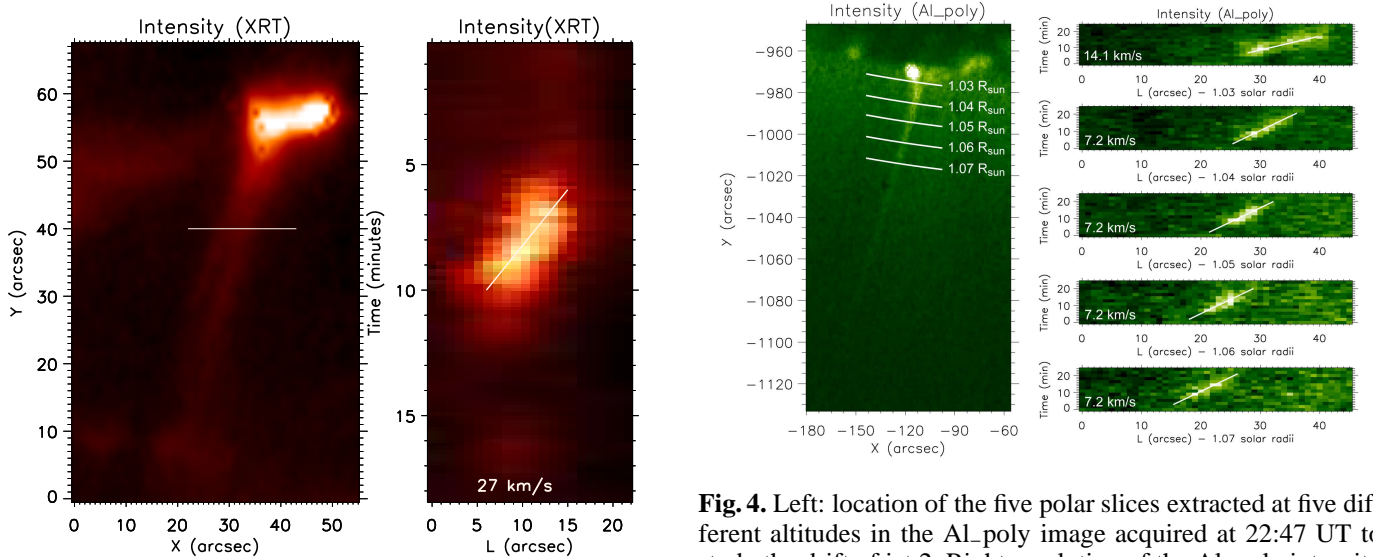}}
\caption{Left panel: the solid white line indicates the location selected to compute the temporal evolution of the XRT intensity across jet~1, which is shown in the right panel. }
\label{js1}
\end{figure}

\begin{figure}
\resizebox{\hsize}{!}{\includegraphics{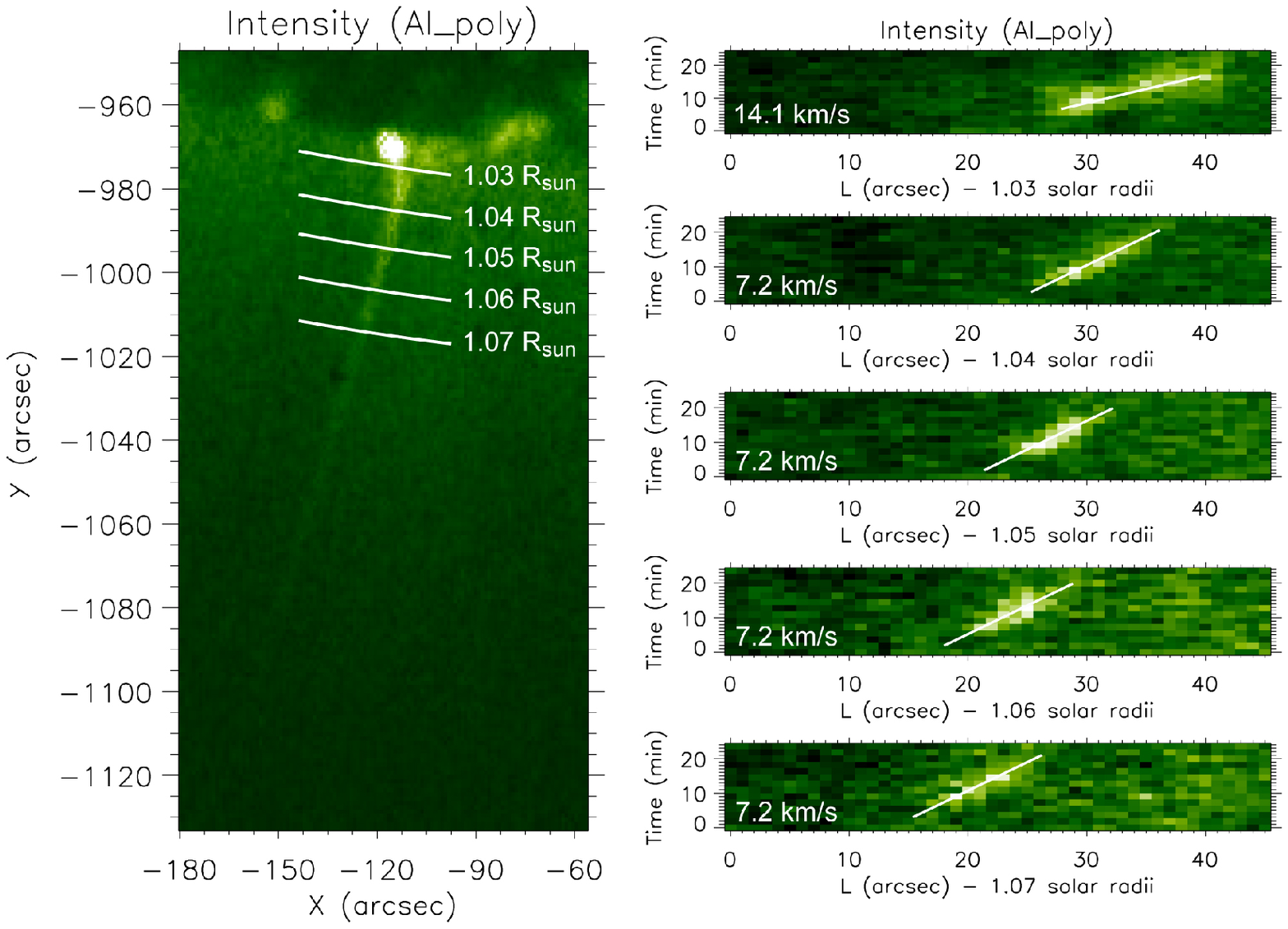}}
\caption{Left: location of the five polar slices extracted at five different altitudes in the Al$_{-}$poly image acquired at 22:47 UT to study the drift of jet~2. Right: evolution of the Al$_{-}$poly intensity along the five slices as shown in the left panel at five different altitudes.}
\label{drift}
\end{figure}

The jet observed on April 15, 2007 was associated with an X-ray bright point (XBP) (Fig.~\ref{context}). From the image sequence (movie~1)  we observe multiple jets originating from the XBP. During the eruption studied here  a significant lateral motion in the direction perpendicular to the jet axis is observed, which we refer to here as drift motion. The left panel of Figure~\ref{js1} shows the XRT jet, and the horizontal white line marks the location along which we  extracted the observed intensities frame by frame perpendicular to the jet (Figure~\ref{js1},  right panel). By measuring the slope, we calculated  a drift of $\approx$10 arcsecs within 5 minutes, resulting in a  speed of $\approx$ 27 km s$^{-1}$.
For the second jet reported here we measured a lower lateral drift speed: the drift was measured by extracting intensity slices along curved lines, centered on the Sun, at five different altitudes above the limb and centered on the average jet latitude (see Fig.~\ref{drift}). The jet drifted at any time at all altitudes by $\approx$ 11 arcsecs over $\approx$ 18 minutes (see Fig.~\ref{drift}), hence with an average drift speed of $\approx$ (7 $\pm$1) km~s$^{-1}$. Only at the base of the jet we measured a drift speed higher by about a factor 2: the interpretation for this result is provided below.


\subsection{Lifetime and expansion speed} 
The April 15, 2007 XRT image sequence (movie~1) shows in an interesting way the occurrence of multiple jets that originate from an XBP. The jet observed around 13:12 UT is the brightest and for this reason was studied in detail. From XRT images we find that the lifetime of the jet used to determine the expansion speed is approximately 5 minutes. The jet-expansion speed has been studied statistically by many authors \citep{1996PASJ...48..123S, 1996ApJ...464.1016C,2007PASJ...59S.771S,1992PASJ...44L.173S}.  \cite{1996PASJ...48..123S}  studied 100 X-ray jets observed with the Soft X-ray Telescope (SXT) onboard Yohkoh and estimated jet velocities on the order of 10-1000~km~s$^{-1}$ with an average velocity of about 200~km~s$^{-1}$. \cite{2007PASJ...59S.771S} extensively studied polar jet formation and evolution with instruments onboard Hinode and SOHO and reported an average speed of 160~km~s$^{-1}$.

\begin{figure}
\resizebox{\hsize}{!}{\includegraphics{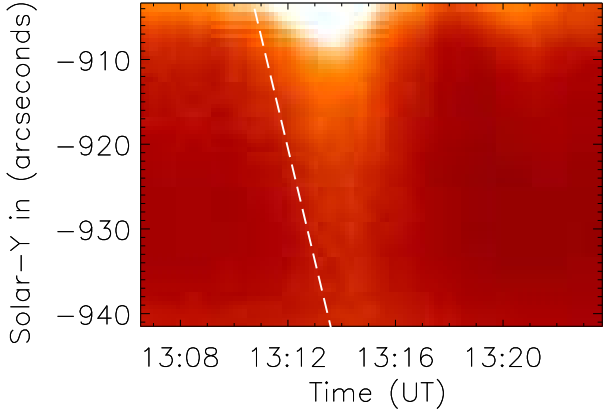}}
\caption{Stack plot for X-ray data relative to jet~1. It shows the variation of the length of the jet brightness averaged over its width as a function of time. The dashed white line provides a measurement the jet speed.}
\label{xrtstackplot}
\end{figure}

We calculated the speed of the jet using different methods. One of the ways of finding the outward speed of the jet is the stack-plot method \citep{2007PASJ...59S.771S}. This method was first applied to X-ray images. We considered an image or box consisting of 37 rows and 24 columns (rectangular box in Fig.~\ref{context}). The box was considered such that it contained the whole length of the jet. Care should be taken so that there are no significant contributions to the intensity other than the jet. We averaged the 24 elements of the row to obtain the width-averaged brightness at all distances from the base row. Thus, for a given instant in time, we obtained a single column to represent the whole image. Plotting these columns for every instant in time, we effectively produced a brightness plot with position and time (Fig.~\ref{xrtstackplot}). This plot is referred to as the stack plot. To measure the velocity of the jet, we fit a line to the left edge of the bright region. The slope of the fitted line  gives an outward velocity of the jet of $\approx$ 158 km s$^{-1}$. The error in this method is estimated as $ \pm $ 35 km s$^{-1}$.

\begin{figure}
\resizebox{\hsize}{!}{\includegraphics{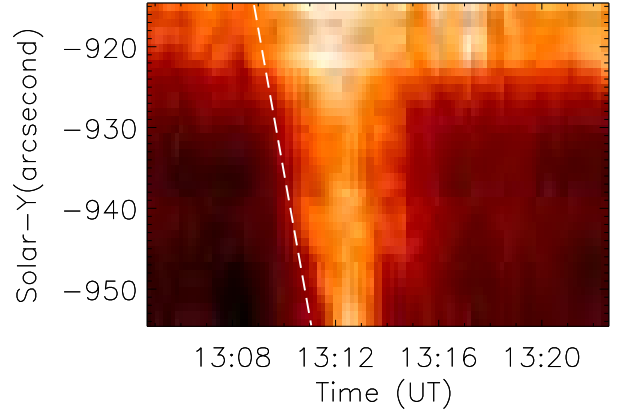}}
\caption{Stack plot for the EIS data. It shows the brightness variation averaged over its width as a function of time corresponding to jet~1. The dashed white line provides a measurement of the jet speed.}
\label{eisstackplot}
\end{figure}

We applied the stack-plot method to EIS slot images. This method has not been applied to the EIS slot images in the past. For EIS images a box size 40 rows and 22 columns was used  (rectangular box in Fig.~\ref{context}).  The resulting stack plot is shown in  Figure~\ref{eisstackplot}. The outward velocity of the jet from EIS data is approximately 172 km s$^{-1}$ . The error in this method is estimated as $ \pm $ 38 km s$^{-1}$.

\begin{figure}
\resizebox{\hsize}{!}{\includegraphics{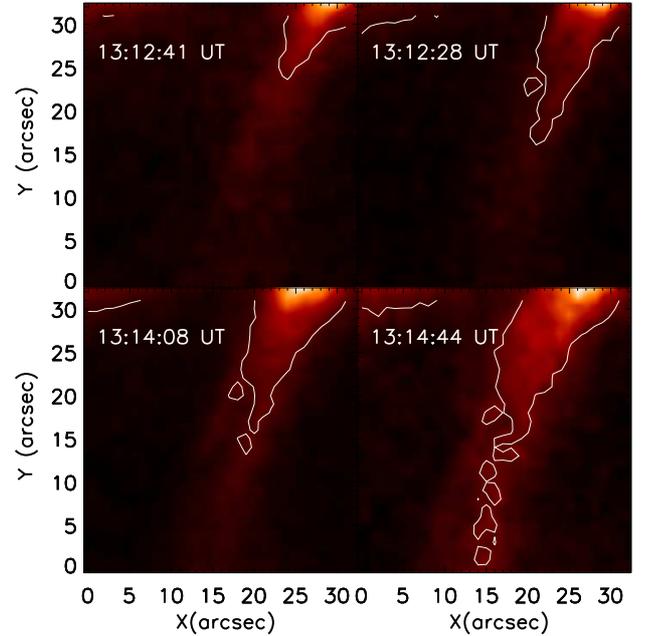}}
\caption{XRT snapshots corresponding to jet~1 and different times,  (a) initiation of the plasma flow, (b) after 45 seconds from the start of the flow, (c) after 90 seconds from the start of the flow, (d) when the jet reaches the end of the box.}
\label{contourmethod}
\end{figure}

We also estimated the jet velocity from the XRT contour images. These images or frames are used to follow the plasma flow with time. Note that at 13:12:41 UT in frame (a) of Figure~\ref{contourmethod} the plasma front or jet front is at position (X,Y) equal to (23.8$\arcsec$, 23.7$\arcsec$). The plasma front reaches (14.1$\arcsec$, 0.2$\arcsec$) at time 13:14:44 UT (frame (d) in Fig.~\ref{contourmethod}). Thus the jet moves over a distance of  $\approx 1.9 \times 10^4$ km within 123 seconds. This method yields a velocity  of the jet of $\approx$ 154~km~s$^{-1}$ with an error of $ \pm $ 40 km s$^{-1}$.

We also estimated the speed of the jet from the light curves. In this method we selected boxes of typical dimensions along the length of the jet and averaged the counts in the box to produce light curves. The boxes were selected such  that they are almost equally spaced. The distance between the boxes was measured, and the time it takes  to travel that distance was found from the light curve. The average velocity measured with this method is 200~km~s$^{-1}$ with an error of $\pm$ 40~km~s$^{-1}$. 

\begin{figure}
\resizebox{\hsize}{!}{\includegraphics{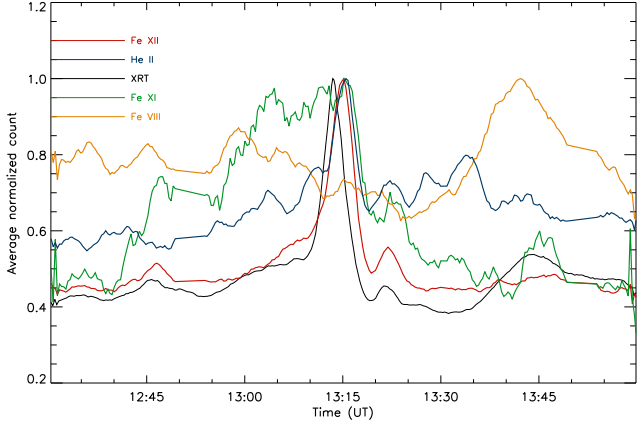}}
\caption{Light curves of XRT and different lines in EIS taken along the length of the jet~1.  The small box (see Fig.~ \ref{context}) taken for light curve analysis corresponds to the upper body of the jet and it is far from the base of the jet.}
\label{lightcurves}
\end{figure}

We inspected the light curves of the different lines/bandpasses in EIS and XRT and compared them with a corresponding region shown as a square box in Figure~\ref{context}. The different ion species contributing to the emission, their formation temperature, and the central wavelength of the band are given in Table \ref{tab1}. EIS data included eight lines, but we  selected four comparatively cleaner lines to plot the light curves. The lower-temperature lines of He~{\sc ii} and Fe~{\sc viii} and the higher-temperature lines Fe~{\sc xi} and Fe~{\sc xii}  were selected to see the response of the atmosphere to the passage of the jet. The light curves in Figure~\ref{lightcurves} in the low-temperature lines (e.g. He~{\sc ii} and Fe~{\sc viii}) show a pronounced enhancement in intensity after the jet peak, while light curves in higher-temperature lines (e.g. Fe~{\sc xii}) are more concentrated at the jet peak. This suggests that either the cool plasma returns to the solar surface, or that we have another jet that is seen only in low-temperature lines. Since the velocity of the jet does not exceed the escape velocity from the Sun, we expect that the response in the low-temperature lines is due to jet material that falls back.
\begin{figure}
\resizebox{\hsize}{!}{\includegraphics{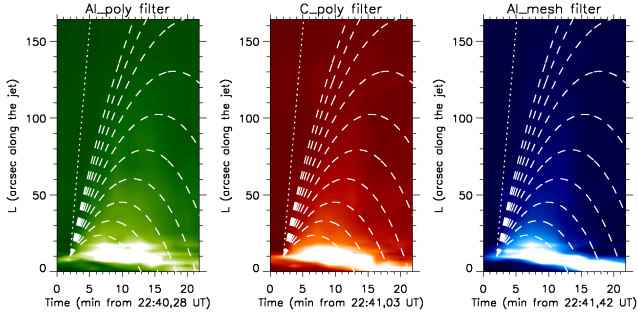}}
\caption{Stack plot for X-ray data relative to the second jet observed during the same time interval with Al$_{-}$poly (left panel), C$_{-}$poly (middle panel) and Al$_{-}$mesh (right panel) filters. Superposed curves show the ballistic trajectories for different initial speeds smaller than the escape speed (dashed lines, see text) and for initial speed equal to the escape speed (dotted line).}
\label{xrtstackplot2}
\end{figure}
\begin{table}
\caption{EIS emission lines.}
\label{tab1}
\centering
\begin{tabular}{c c c}
\hline\hline
central wavelength   & $T_{e}$  &  contributing \\
$ \lambda$  (\AA)   &    MK          &  Ion species \\
\hline
256.32             & 0.079               & \ion{He}{ii}, \ion{Si}{x}, \ion{Fe}{xii}, \ion{Fe}{xiii}, \ion{Ni}{xvi},\ion{S}{xiii}              \\
 195.12            & 1.26             & \ion{Fe}{xii}, \ion{Fe}{viii}, \ion{Ni}{xvi}, \ion{Ni}{xv} \\
 185.21            & 0.40               &   \ion{Fe}{viii}, \ion{Ni}{xvi}   \\
 180.40            & 1.0               & \ion{Fe}{xi}, \ion{Fe}{x} \\

\hline
\end{tabular}
\end{table}
The expansion velocity of the second jet reported here, observed on July 1, 2008, was also studied with the same technique as for XRT data acquired with the three different available filters. The resulting jet stack-plots are shown in Figure~\ref{xrtstackplot2} for the Al$_{-}$poly (left), C$_{-}$poly (middle), and Al$_{-}$mesh (right) filters. Superposed to the jet stack-plots we also show the expected altitude-versus-time curves for ballistic motions of material ejected with initial velocities ranging between 70 km s$^{-1}$ and 250 km s$^{-1}$ (dashed lines) in steps of 20 km s$^{-1}$. These curves were computed taking into account the observed inclination of the jet axis with respect to the radial direction and the variation of the gravitational acceleration with altitude. A comparison between the jet stack-plots and the ballistic trajectories in this figure leads us to conclude that the jet expands with a velocity not higher than $(250 \pm 50)$ km s$^{-1}$, again much lower than the escape velocity from the Sun. More importantly, the jet stack-plots in Figure~\ref{xrtstackplot2} show a progressive rising in altitude, a peak around 22:52 UT, followed by a progressive decay. Hence, the stack-plots have a clear symmetry about the $x-$axis (i.e. about the time evolution): this shape of the plots, visible also in Figure~\ref{xrtstackplot} for the first jet reported here, does not agree with the idea of a packet of plasma ejected and simply travel that is outward: we return to this point below.

\begin{figure}
\begin{center}
\includegraphics[width=9cm]{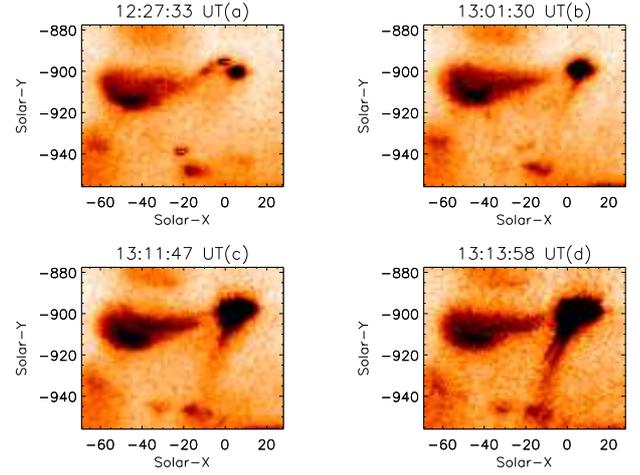}
\caption{Fine structure of an XBP corresponding to jet~1 in the southern polar coronal hole as observed by XRT (Al poly filter). The morphology of the XBP and the different jet phenomenon in the micro sigmoid are shown in the figure (see online movie~1).}
\label{p}
\end{center}
\end{figure}
\subsection{Evolution of the sigmoid structure}
A micro-sigmoid region is often associated with the eruption of multiple jets. Some of the jets observed in April 2007 are double-stranded. Recurrence of the jets is associated with recurring reconnections in the bright-point region.  A closed-loop  structure interacting with the coronal open magnetic fields gives rise to multiple reconnection. The bright-point morphology  associated with the jet  changes continuously. \cite{1996ApJ...464L.199R} observed a transient S-shaped brightening prior to the eruption of coronal mass ejections (CMEs). These S-shaped brightenings are complex helical flux-ropes and the erupting flux rope expands to form the CME. A coronal sigmoid can be transient as well as long lasting. \cite{1999GeoRL..26..627C} suggested that active regions with sigmoidal X-ray morphologies were 68 percent more likely to erupt than non-sigmoidal active regions. Extensive observations were reported in \cite{2010ApJ...718..981R} about the micro sigmoid structure, evolution, and their relation with the formation of the associated jets. 
\begin{figure}
\resizebox{\hsize}{!}{\includegraphics{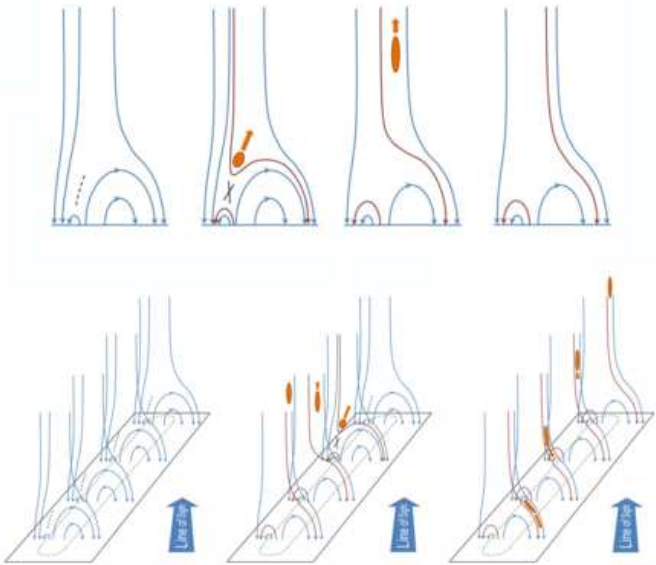}}
\caption{Top: sequence of cartoons illustrating the occurrence of a single reconnection event in one of the pseudo-streamer-like arcades considered here (blue field lines) leading to a classical standard jet, see Figure~1~of \cite{MOOetal2010}. The interchange reconnection occurs in the neutral current sheet (dashed line), leading to the formation of newly open and closed (red) field lines and to the ejection of a plasma packet along the open field lines.. Bottom: proposed interpretation of the observed jet observational properties. In particular, the drift is ascribed to many small-scale reconnections occur sequentially along a spine inclined with respect to the line of sight (see text).}
\label{cartoon}
\end{figure}
In this study for the April 2007 case, we see a micro-sigmoid of size $70\arcsec \times 30 \arcsec$ in XRT images. Before and after the occurrence of the jet, the bright point showed many changes in morphology. In  Figure~\ref{p} we show some snapshots of the jet evolution and the associated bright point (see also  online movie~1). At about 12:27 UT (Fig.~\ref{p}~a) the intensity of the bright point is enhanced and the sigmoidal shape of the loop structure is clearly visible. We assume that this corresponds to the energy deposition to the bright point, and it is followed by an increase in the size of the bright point and eruptions that lead to multiple jets. At about 12:45 UT we see the first lambda-type jet that originates from the right side of the bright point (see movie~1). This jet has a low intensity and short lifetime.  After approximately 15 minutes at about 13:01 UT (Fig~\ref{p}~b), we see a second lambda-type jet that originates from the right side of the bright point. By this time the bright point shows a loop-like structure. At about 13:10 UT (see movie~1) one can clearly observe two jets simultaneously side by side. One of the jets, as previously observed, originates from the right side of the loop structure, while the other originates from the left side of the loop structure. Both seem to be lambda-type jets. The jets seen so far were all faint and can be seen in only two to three XRT images and thus could not be used for velocity measurements.  At about 13:12 UT (Fig~ \ref{p}~c) we observe a jet that  originates from the right end of the bright-point loop structure, which develops into  a bigger structure with time, and at about 13:13 UT (Fig~ \ref{p}~d) we see that the jet  has two strands. This jet was used to study drifts, as reported in the previous subsections. More evidence of the multiple jet features is seen at  13:20 UT (see movie~1). The subsequent images (see movie~1) show that the loop structure completely diffuses, which is followed by the complete disappearance of the bright point. One should also note that the size of the brightening structure, namely the bright  point, is similar to the lateral drift, which  suggests that multiple reconnection or a stepping reconnection scenario is acting here. For jet~2 we also found that the length of the  lateral drift motion is similar to the length of the associated brightening at the jet's base. 

\cite{2010ApJ...718..981R} have studied several jet events in the coronal holes and found that they erupt from small-scale S-shaped bright regions. They furthermore suggested that coronal micro-sigmoids may well be the progenitors of coronal jets. Our case support this scenario as well. 

\section{Jet fountain model}
\label{model}
\subsection{Current stage of polar jet simulation}
A 3D MHD model of polar jets was proposed by \cite{2009ApJ...691...61P}, where a jet is produced starting from an axisymmetric configuration of magnetic fields, to which an axisymmetrical twisting motion at the photospheric boundary is applied, accumulating free magnetic energy and injecting magnetic helicity into the system. Eventually, a burst of magnetic reconnection occurs, creating new twisted open field lines where Alfv\'en waves propagate, which accelerates the plasma via pressure gradients and creates a collimated plasma ejection. During the reconnection process, the vertical spine migrates and, viewed from one side, the jet is observed to drift. This interpretation of the jet drift implies a drift velocity ($v_{d}$) of ``about one fifth to one tenth of the upward velocities found in the jet itself'', an outflow upward velocity ($v_{out}$) on the order of the local Alfv\'en speed, and a drift motion occurring over ``a length scale of the order of the width of the closed magnetic field'' \citep[$v_{out}$; see][]{2009ApJ...691...61P}, hence over a length similar in size with the jet width. The reference estimated value of the local Alfv\'en speed for this simulation is $v_A = 1400$~km~s~$^{-1}$.

These conditions seems to be not met for the July 2008 event and are only marginally met for the April 2007 event described above. In particular, the July 2008 jet expands with an outflow velocity on the order of $v_{out} \sim 200$~km~s~$^{-1}$, a factor $\sim 30$ higher than the observed drift speed. In the \cite{2009ApJ...691...61P} model the drift speed is expected to be $v_{d} \sim 0.2 \, v_A$, where $v_A$ is the local Alfv\'en speed: with the measured drift velocities $v_d \sim 27$ km s$^{-1}$ and $v_d \sim 7$ km s$^{-1}$ this implies $v_A \sim 130$ km s$^{-1}$ and $v_A \sim 35$ km s$^{-1}$, respectively, while in a polar coronal hole $v_A$ is expected to be on the order of $\sim 1000$ km s$^{-1}$  \citep{2005ApJS..156..265C}. On the other hand, in the \cite{2009ApJ...691...61P} model the ejected plasma is expected to expand with a very high outflow velocity of the order of the local $v_A$, but again the observed outflow velocities $v_{out}$, on the order of 171 km~s $^{-1}$ and 250 km~s $^{-1}$, seem to be far too low compared with the expected values of $v_A$ mentioned above. Moreover, the observed drift of $\sim 10$ arcsecs covers an area about twice as large as the jet width, which does not change throughout the event and hence higher than envisaged by the \cite{2009ApJ...691...61P} model. Even if the first discrepancy between the lower-than-expected drift velocity and the outflow speed can be explained by projection effects, the same explanation would imply a higher-than-observed real drift, which would increase the second discrepancy between the drift size and the jet width even more. 

It is also very important to notice that a single violent burst reconnection event seems unable to explain the very long lifetime ($\sim 20$ minutes) of the observed jet, its constant drift motion at different altitudes, and its constant transverse speed. In particular, because the point of reconnection between closed and open field lines is migrating and at the same time the open field lines are expected to be parallel (in first approximation) in the small altitude range we consider (0.04 solar radii), we expect the drift velocity to be constant at different altitudes, as observed. The higher drift speed we measured at the base of the jet could be much closer to the real drift speed of the reconnection point, ejecting plasma that is channeled along open field lines. Furthermore, as we mentioned before, the observed jet stack-plots show a clear symmetry around the time axis, suggesting that some material is not only ejected, but is also flowing back to the Sun. Nevertheless, given the available spatial and temporal resolutions, XRT data show no evidence for single blobs or plasma packets traveling outward or inward. Hence, if many small-scale reconnection events are occurring during the jet event, the size of the blobs should be smaller than the instrumental resolution, and the ejection rate should be faster than the observation rate.  \cite{MOOetal2010} found that X-ray jets in the polar coronal holes can be classified based on their morphology into two different types, standard jets and blow-out jets. For standard jets, the spire is a single spike with the typical inverted-Y shape, and there is a bright point on the edge of the base, usually at or just outside one of the footpoints  of the spire. For blow-out jets,  the spire is more complex during the growth phase than in standard jets. They have the additional emission of a cool plasma component that is visible in spectral lines that typically correspond to chromospheric and transition region emission. The other difference is blow-out X-ray jets show a strong X-ray brightening in the core of the base arch. The two jets observed in our study are standard jets. The spire of the jets mostly has a single spike and an inverted-Y shape. Following the distinction proposed by \cite{MOOetal2010}, the discrepancies may be related to the fact that  the  model by \cite{2009ApJ...691...61P}  correctly describes the occurrence of so-called blow-out jets, but  may not describe the standard jets correctly. To investigate the possibility that these observational discrepancies could be the signature of a different origin for the occurrence of polar jets reported here, we developed a toymodel simply aimed at reproducing the observed characteristics of polar jets considering a different scenario. 

\subsection{Proposed polar jet toy-model}

In what follows we assumed that the observed jets are generated by a sequence of single reconnection events where single unresolved packets of plasma, smaller than the instrumental spatial resolution, are ejected along open field lines, then expand and fall back along the same path, following a simple ballistic motion. The top part of Figure~\ref{cartoon} shows the sequence of events (from left to right) in the single-arcade system: as magnetic reconnection is initiated, a packet of plasma is ejected outward and is then deflected and channeled by the open fieldlines. The bottom part of this figure shows the geometrical 3D configuration we considered: a system of asymmetric pseudo-streamer-like arcades is distributed along a straight neutral line, inclined with respect to the direction of the line of sight (bottom left). As reconnection in one single point of the neutral spine is initiated, inflows from the surrounding corona toward the reconnection point trigger more reconnections in the nearby locations along the same spine (bottom middle). Hence, the reconnection propagates along the spine, ejecting plasma packets outward at different times in different locations; eventually the plasma packets expand and flow back on to the Sun (bottom right).

\begin{figure}
\resizebox{\hsize}{!}{\includegraphics{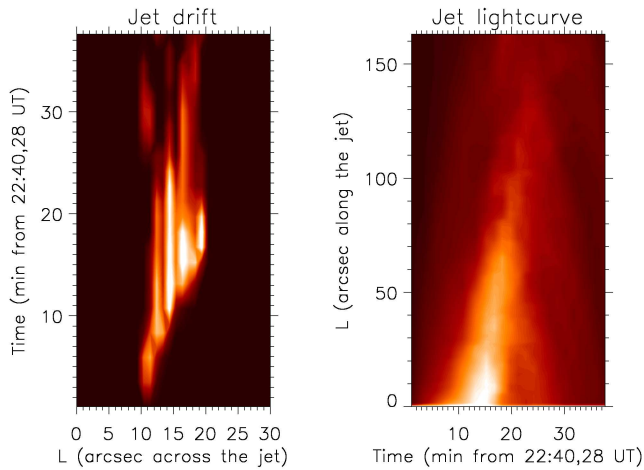}}
\caption{Left: evolution of the simulated intensities averaged in a direction perpendicular to the jet, to allow a direct comparison with drift observations (see Fig.~\ref{js1}, right panel, and Fig.~\ref{drift}, right panels). Right: evolution of the simulated intensities averaged in a direction parallel to the jet to allow a direct comparison with stack-plot observations (see Fig.~\ref{xrtstackplot} and panels in Fig.~\ref{xrtstackplot2}).}
\label{simulation2}
\end{figure}

In particular, to reproduce XRT observations relative to the July 1, 2008 jet, we assumed that for each reconnection event a cylindrical plasma packet with a width on the order of $1.81 \times 10^7$ cm (equivalent to a projected size of 0.25 arcsecs, 1/4 of a single XRT pixel), a mass of $10^3$~kg, and a density of $10^{10}$ cm$^{-3}$ is ejected with an initial velocity following a Gaussian distribution of random outflow velocities with an average speed of 200 km~s$^{-1}$ and $\sigma = 116$ km~s$^{-1}$. The path followed by each plasma packet has an inclination of 21$^\circ$ with respect to the radial direction for the specific case analyzed here.  Unresolved plasma packets expand along magnetic-field lines, maintaining a constant density during their evolution: given the small altitude interval considered here (about 0.15 solar radii), a possible superadial expansion of magnetic fluxtubes was neglected. Single reconnection events occur at a different spatial location following a straight neutral line with an inclination of $30^\circ$ with respect to the line of sight (see Fig.~\ref{cartoon}). The length of the spine involved in these subsequent reconnections is $1.45 \times 10^9$ cm (equivalent to 20 arcsecs). The unknown length of the spine $L$, where the multiple reconnections occur and its inclination angle $\theta_{LOS}$ with respect to the line of sight, are simply related to the observed drift $d$ to be reproduced by $d = L \sin \theta_{LOS}$. The total duration of the energy release process (i.e. ejection of plasma packets) was set equal to 15 minutes; the subsequent evolution is simply due to the plasma propagation. The simulation considers the projected ballistic motions of the plasma packets ejected at different locations along the 3D spine, performs the integration along the line of sight for each time step  (by assuming the intensity to be proportional to the square of the density), and then down-grades the resulting 2D images to the XRT instrumental resolution and pixel size of about 1 arcsec; no thermodynamic evolution of the plasma is taken into account so far (see also the online jet animation, movie~3: \url{ftp://ftp.iiap.res.in/chandra/xrt-jet/movie3.gif} \label{m3}).
\begin{figure*}
\resizebox{\hsize}{!}{\includegraphics{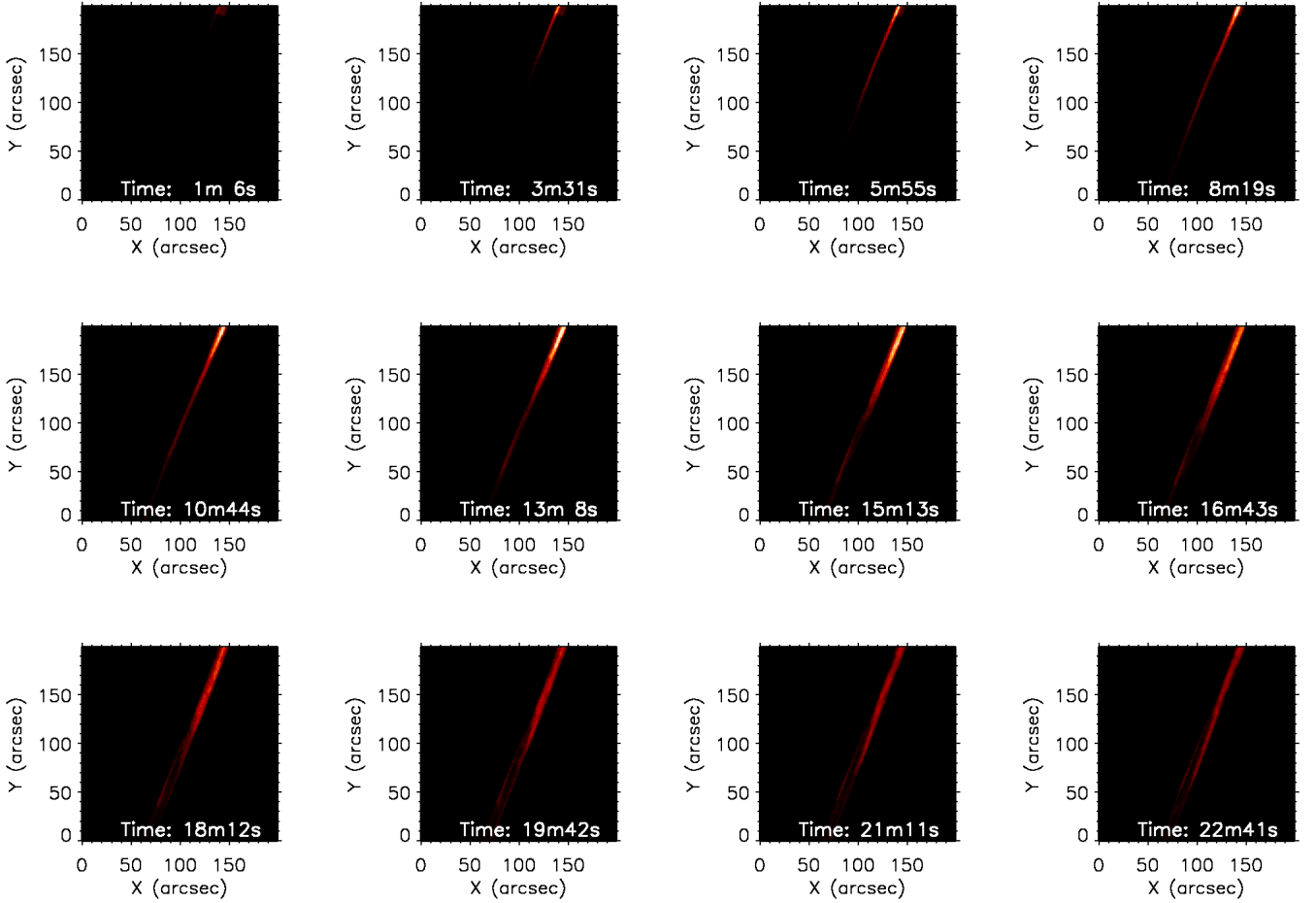}}
\caption{Sequence of jet-simulated images as obtained with the fountain-model proposed here, assumes multiple small-scale reconnections where many unresolved blobs are ejected along open field lines following a simple ballistic motion (see also online movie~3).}
\label{simulation}
\end{figure*}
Simulation results are plotted as a sequence of 12 snapshots (taken with the same time-resolution as the XRT observations) in Figure~\ref{simulation}. This sequence shows that this toymodel is able to reproduce at the same time 1) the observed low drift velocity of the jet, related with the low projected inclination of the neutral line with respect to the line-of-sight; 2) the amount of the observed drift, which much larger than the jet width, because of drift of the reconnection point along an inclined neutral line; and 3) the long persistence of the jet, due to the occurrence of many unresolved small-scale blobs that are ejected at different times with a Gaussian distribution of the initial velocities. The model also reproduces the jet stack-plots well. The simulated jet-drift curve and stack plots are also shown in Figure~\ref{simulation2} for direct comparison with the observations. The simulation results clearly show that  the jet stack-plots not have the typical shape expected when an amount of plasma is really ejected outward in a single reconnection event, but instead shows a parabolic envelope of a jet becoming more and more elongated in the earlier phase of its evolution that then progressively decays in the later phase. The model also produces single jet images where the plasma appears to be spatially structured within the jet, an observational property that is harder to explain with a single reconnection event. The total kinetic energy $E_{kin} = 1/2 \rho v_{out}^2$ that is released during the whole process (i.e. summing over all ejected plasma packets) is on the order of $E_{kin} \simeq 3.5 \times 10^4$ erg cm$^{-3}$: obviously, this is only a lower limit to the total energy injected because other energy sources (such as thermal energy, waves) have not been taken into account here.
%
\section{Conclusions}
We presented the analysis of two jets observed in April 2007 and July 2008 in a polar coronal hole by Hinode/XRT. Of all the jets observed, the jets at 13:12 UT on April 15 2007 and  at 22:50 UT on July 1 2008 are the brightest and are visible for approximately 5-10 minutes. These jets were studied in detail: we found no evidence of helical motions in these events, but detected a significant shift of the jet position in a direction normal to the jet axis, with a drift velocity of about 27 km~s$^{-1}$ and 7 km~s$^{-1}$, respectively. The jet expansion velocities were calculated with the stack-plot method and other methods. For the first time, we applied the stack-plot method to the EIS slot images. The average apparent velocities of the first and second jets derived with different techniques are 171 km~s $^{-1}$  and 250 km~s $^{-1}$, both much lower than the escape velocity from the Sun. The evolution of the jet intensities in XRT data showed that both jets are observed as an expansion of material, followed by a subsequent contraction. Hence, resulting XRT stack plots are approximately symmetric about the time when the jet brightness peaks. Moreover, light curves derived with Hinode/EIS in low-temperature lines show a pronounced enhancement in intensity after the jet intensity peak. These two observational facts suggest that the ejected plasma is most likely flowing back no to the Sun in both events. In agreement with this conclusion, \cite{2012A&A...538A..34F} recently demonstrated that the kinematic trajectories of solar jets could be successfully fitted with a simple  ballistic model. \cite{2012A&A...538A..34F} applied a kinetic analysis to ejected particles reproducing, a large polar jet observed by EUVI and COR1 by neglecting the magnetic mirror force and particle collisions, thus simply considering the effect of gravity, as we also did. Hence, results described here extend the conclusions by \cite{2012A&A...538A..34F}, showing that even smaller-scale polar jets observed only in the field of view of EUV imagers can be well reproduced by a simple ballistic model.

Nevertheless, the simple fact that the jet expansion velocities are found to be much lower than the escape velocity is obviously not a demonstration that the plasma will eventually fall back onto the Sun. In fact, if the plasma is accelerated by magnetic reconnection between closed field lines in the bright points and the nearby open field lines, even if the stronger acceleration will occur at the reconnection time and at the reconnection point, an additional acceleration could be provided higher up by the tension of the newly reconnected field lines. Alternatively, an additional acceleration could also be provided by the higher plasma pressure expected at the base of the jet, where plasma heating due to reconnection will occur. For instance, \cite{2012SoPh..tmp..310M} recently demonstrated that the jet expansion is caused by a combination of the downward gravitational force and an upward driving force that weakens the decelerating effect of the solar gravitational field. For an EUV jet which occurred in an active region, \cite{2012ApJ...759...15M} recently concluded that the cool plasma (observed in He~{\sc ii} and Fe~{\sc viii} lines) was ejected by magnetic acceleration due to magnetic reconnection, while the hot plasma (observed in Fe~{\sc xii} and Fe~{\sc xvi} lines) was ejected by thermal acceleration similar to the chromospheric evaporation.

Whatever acceleration process is considered for the ejection of plasma, the amount of plasma that eventually escapes from the Sun and contributes to the solar wind mass flow and energy is at present not clearly quantified. Images provided by band-pass filters (such as those provided by Trace, Hinode/XRT, or SDO/AIA) can show plasma appearing and/or disappearing, depending also on variations of plasma temperatures that move the emission in or out of the filter temperature response. In any case, if the plasma is eventually ejected and the jet occurs close to the limb, a signature should be observed also in white-light coronagraphs: data from the STEREO Heliospheric Imagers, for instance, have proven to be useful to study transient small-scale (down to $\sim 100$ km) structures propagating in the solar wind \citep{2008GeoRL..3524104D,2009GeoRL..36.2102D,2010SoPh..265..207D}. In the lower corona, small-scale blobs observed by STEREO/COR1 have been reported \citep{2009ApJ...701.1906J}, and small-scale density structures ($\sim 100-1000$ Mm) are currently also studied with COR1 data \citep{2013arXiv1302.3382T}. Many EUV jets have been identified higher up in COR1 images, as reported by \citet{2009SoPh..259...87N}, but only a fraction of 73-75\% of COR1 jets was found to be associated with a clear EUV jet, as reported by \citet{2010SoPh..264..365P}.For the two events reported here we investigated the possible expansion of ejected plasma into the lower corona by looking at white-light images acquired by STEREO/COR1 and SOHO/LASCO coronagraphs one hour after the event. With these data, standard running-difference frames were built, to enhance possible weak intensity variations due to the jet expansion: however, we found no evidence of these jets in the white-light images. From the theoretical point of view, this lack of detection only suggests the possibility that the jet plasma was not escaping from the Sun, but it cannot be considered as true demonstration of this, given the small extension and short duration of these features and their density decay due to the expansion in the corona. Moreover, we checked for the presence of these two jets in EUVI data and found none: this could be due to the lower resolution, the lower cadence, or both, with respect to XRT data or it could be because the XRT jets were not observed by SECCHI because of the temperature of the observed transients. The XRT sensitivity decreases rapidly below temperatures of a few MK, which means that a very hot jet may not have a strong enough signal to be detected by EUV channels of SECCHI. In any case, on April 15, 2007 (first jet) the angle between STEREO spacecraft was about 4.3 degree, hence the source region will be observed by EUVI telescopes approximately along the same line of sight as XRT data. In contrast, on July 1, 2008 (second jet) the angle between the STEREO spacecraft was about 58.7 deg, but the jet occurred very close to the pole, hence EUVI data will be affected by the same projection effects that affect the XRT data, which precludes studying the source region.

The contribution to the solar wind could be also related to the type of jet: for instance, \citet{2013SW13} recently showed that of the two categories of jets (blow-out and standard) proposed by \citet{MOOetal2010}, only the blow-out jets might contribute to the solar wind mass flux. Hence, to further support our conclusion that the jet plasma flows back onto the Sun, we also proposed a simple qualitative toymodel that mimics the observations. In this model the observed behavior was interpreted within a scenario where reconnection progressively shifts along a magnetic structure, leading to the sequential appearance of jets of about the same size and physical characteristics. Results from this fountain model showed that a simple ballistic motion of multiple unresolved plasma blobs is able to explain at the same time the observed jet drift, the life time, and stack plot and hence all the basic observational properties described here. Therefore these results suggest that  jets similar to those reported here can provide no significant contribution to the fast solar wind mass flux that emerges from polar coronal holes, even if a possible contribution to the heating of the coronal base cannot be excluded.

\begin{acknowledgements}
 This work was supported by the Indo-German DST-DAAD joint project D/07/03045. Hinode is a Japanese mission developed and launched by ISAS/JAXA, with NAOJ as domestic partner and NASA and STFC (UK) as international partners.  It is operated by these agencies in co-operation with ESA and NSC (Norway). The authors also thank G. Poletto and A. C. Sterling for providing XRT data acquired during the July 2008 campaign. We would like to thank the referee for his/her valuable comments, which have enabled us to improve the presentation.
\end{acknowledgements}

\bibliographystyle{aa.bst}
\bibliography{chandu.bib}

\Online

% ...
%
% ...

\end{document}